# Dynamical van der Waals atom –surface interaction

G.V.Dedkov[1] and A.A.Kyasov

Nanoscale Physics Group, Kabardino –Balkarian State University, Nalchik, 360004

Russian Federation

We obtained new nonrelativistic expression for the dynamical van der Waals atom –surface interaction energy of very convenient form for different applications. It is shown that classical result (Ferrell and Ritchie, 1980) holds only for a very slowly moving atom. In general case, the van der Waals atom –surface interaction energy manifests strong nonlinear dependence on velocity and distance. In close vicinity of metal and dielectric surfaces and velocities from 1 to 10 Bohr units the dynamical van der Waals potential proves to be significantly lower that in static case and goes to the static values with increasing distance and (or) decreasing velocity

## 1.Introduction

Since 1980's, it has been recognized that dynamical correction to the van der Waals interaction potential between a moving ground state atom and a surface, to leading order, is proportional to the squared velocity, $V^2$ [1,2]. This basic result, obtained by Ferrell and Ritchie [1], has been later generalized by Annett and Echenique [2] with account of the surface plasmon dispersion. Recently, Barton [3] has reproduced this result using the second order perturbation theory. Also, a problem of dynamic corrections to the van der Waals energy has been studied in our papers [4] using nonrelativistic limit of fluctuation electromagnetic theory and, more recently [5], in relativistic treatment.

The aim of this paper is to summarize and to develop further our general results [4,5] with no account both retardation and relativistic effects. Moreover, throughout we assume the temperature to be zero and an atom in the ground state. A noticeable point of the present analysis appears to be a more complex dependence of the van der Waals potential on the velocity and distance. We show that formula obtained by Ferrell and Ritchie [1] holds only for a very slowly moving atom. At higher enough still nonrelativistic velocity of an atom the van der Waals potential becomes significantly lower (by modulus) than at $V = 0$ at the distances to the surface of 0.3 to 1 $nm$. We illustrate this behavior numerically in the case of metastable $He^*$ and ground state *He* atoms with velocities $10^6 \div 10^7 \, m/s$, moving parallel to metal (*Au*) and dielectric ($SiO_2$) surfaces.

---

[1] Corresponding author e-mail: gv_dedkov@mail.ru

## 2. General formulation

At first, we shall summarize different expressions for the van der Waals atom –surface energy, obtained in our papers [4,5]. Thus, in the nonrelativistic and nonretarded approximation, the basic formula for a particle moving with velocity $V$ at a distance $z_0$ above a surface in $x-$ direction (see fig.1), is given by [4]

$$U(z_0,V) = -\frac{\hbar}{\pi^2} \int_0^\infty d\omega \int_0^\infty dk_x \int_0^\infty dk_y \, k \exp(-2kz_0)$$
$$\cdot \left\{ \coth\frac{\omega\hbar}{2k_B T_1} \alpha''(\omega)[\Delta'(\omega - k_x V) + \Delta'(\omega + k_x V)] + \right.$$
$$\left. + \coth\frac{\omega\hbar}{2k_B T_2} \Delta''(\omega)[\alpha'(\omega - k_x V) + \alpha'(\omega + k_x V)] \right\} \quad (1)$$

where $T_1$ and $T_2$ are the temperatures of the particle and surface. One primed and double primed functions $\alpha'(\omega), \alpha''(\omega)$ denote real and imaginary parts of the particle polarizability,

$$\Delta(\omega) = \frac{\varepsilon(\omega) - 1}{\varepsilon(\omega) + 1}, \quad (2)$$

$\varepsilon(\omega)$ denotes bulk dielectric permittivity of half –space, and $\Delta'(\omega)$ and $\Delta''(\omega)$ correspond to real and imaginary parts of $\Delta(\omega)$.

Another form of (1), as adopted in the nonrelativistic and nonretarded case, has been obtained in [5] and is given by (with a trivial transformation from attraction force to the potential of attraction):

$$U(z_0,V) = -\frac{\hbar}{\pi^2} \int_0^\infty d\omega \int_{-\infty}^\infty dk_x \int_{-\infty}^\infty dk_y \, k \exp(-2kz_0)$$
$$\cdot \left[ \Delta''(\omega)\alpha'(\omega + k_x V)\coth\frac{\hbar\omega}{2k_B T_2} + \Delta'(\omega)\alpha''(\omega + k_x V)\coth\frac{\hbar(\omega + k_x V)}{2k_B T_1} \right] \quad (3)$$

Assuming $T_1 = T_2 = 0$ with account of relations

$$\lim_{T_2 \to 0} \coth \frac{\hbar \omega}{2kT_2} = sign(\omega)$$

$$\lim_{T_1 \to 0} \coth \frac{\hbar(\omega + k_x V)}{2kT_1} = sign(\omega + k_x V),$$

eq.(3) takes the form

$$U(z_0, V) = -\frac{\hbar}{\pi^2} \int_0^\infty d\omega \int_0^\infty dk_x \int_0^\infty dk_y\, k \exp(-2kz_0) \cdot$$
$$\left\{ \int_0^\infty d\omega \begin{bmatrix} \Delta''(\omega)\alpha'(\omega + k_x V) + \Delta''(\omega)\alpha'(\omega - k_x V) + \\ \Delta'(\omega)\alpha''(\omega + k_x V) + \Delta'(\omega)\alpha''(\omega - k_x V) \end{bmatrix} - \right.$$
$$\left. - 2\int_0^{k_x V} d\omega \Delta'(\omega)\alpha''(\omega - k_x V) \right\} \quad (4)$$

Just the same formula (4) stems from (1) introducing in the first line of (1) new variables $\omega \pm k_x V = \omega'$ and performing several straightforward transformations.

The first addend in (4) can be simplified further by replacing limits of integration $0 \le k_x, k_y < \infty$ by $-\infty \le k_x, k_y < \infty$ and turning integration contour to imaginary frequency axis. Then, finally, dynamical van der Waals interaction energy proves to be given by

$$U(z_0, V) = -\frac{\hbar}{2\pi^2} \int_{-\infty}^{+\infty} dk_x \int_{-\infty}^{+\infty} dk_y\, k \exp(-2k z_0) \cdot \operatorname{Im}\left[ i \int_0^\infty d\xi\, \Delta(i\xi)\alpha(i\xi + k_x V) \right] +$$
$$+ \frac{2\hbar}{\pi^2} \int_0^\infty dk_x \int_0^\infty dk_y\, k \exp(-2k z) \int_0^{k_x V} d\omega\, \Delta'(\omega)\alpha''(\omega - k_x V) = U^{(0)}(z_0, V) + \Delta U(z_0, V) \quad (5)$$

It is worth noting that eq.(5) does not follow from [1] and until now it has not been presented elsewhere. This is our central result in this paper.

### 3. Limit of small velocities

Using a nondissipative model of metallic half–space,

$$\varepsilon(\omega) = 1 - \omega_p^2/\omega^2, \; \Delta(i\xi) = \frac{\omega_s^2}{\omega_s^2 + \xi^2}, \; \omega_s = \omega_p/\sqrt{2}, \tag{6}$$

and atomic polarizability determined in a single oscillator model,

$$\alpha(i\xi) = \frac{\alpha(0)\omega_0^2}{\omega_0^2 + \xi^2} \tag{7}$$

where $\omega_p$ is the plasma frequency, $\alpha(0)$ is the static value of the dipole polarizability, and $\omega_0$ is the atomic transition frequency, the term $U^{(0)}(z_0,V)$ in (5) simplifies to

$$U^{(0)}(z_0,V) = -\frac{2\hbar\alpha(0)\omega_s^2\omega_0^2}{\pi^2} \int_0^\infty dk_x \int_0^\infty dk_y\, k\exp(-2kz_0) \cdot$$
$$\cdot \int_0^\infty d\xi \frac{(\omega_0^2 + \xi^2 - k_x^2 V^2)}{(\omega_s^2 + \xi^2)[(\omega_0^2 + \xi^2 - k_x^2 V^2)^2 + 4\xi^2 k_x^2 V^2]} \tag{8}$$

The frequency integral in eq.(8) is calculated explicitly and is given by

$$J(a) = \frac{\pi}{2\omega_s(\omega_0 + a)(\omega_s + \omega_0 + a)} +$$
$$+ \frac{\pi a\, sign(\omega_0 - a)(\omega_s + \omega_0 + a + |\omega_0 - a|)}{\omega_s(\omega_0 + a + |\omega_0 - a|)(\omega_0 + a)(\omega_s + \omega_0 + a)(\omega_s + |\omega_0 - a|)}, \; a = k_x V \tag{9}$$

Using the limit of a slowly moving atom, $a = k_x V \ll \omega_0$, one easily finds from (9), to second order in $a^2$:

$$J(a) \approx \frac{\pi}{2\omega_s\omega_0(\omega_s + \omega_0)} \left[1 + \frac{k_x^2 V^2}{(\omega_s + \omega_0)^2}\right] \tag{10}$$

Then, substitution into (8) via (10) leads to

$$U^{(0)}(z_0,V) = -\frac{\hbar\alpha(0)\omega_s\omega_0}{8z_0^3(\omega_s + \omega_0)}\left[1 + \frac{3V^2}{2z_0^2(\omega_s + \omega_0)^2}\right] \tag{11}$$

Eq.(11) exactly coincides with [1] if use is made of a more general quantum expression for the atomic polarizability. Moreover, as will be shown in Section 4, in the case of a slowly moving atom, the term $\Delta U(z_0,V)$ in (5) proves to be exponentially small, so the result by Ferrell and Ritchie holds also for $U(z_0,V)$, as well. However, the general result obtained in [1] (see eqs.(14),(16)) in the case $2\omega_0 z/V \sim 1$ turns out to be essentially different than (8) and (5).

## 4. General case

Now let us consider the case when velocity of an atom is not small, still being nonrelativistic. Substituting (9) in (8) and performing integration over $k_y$ finally yields

$$U^{(0)}(z_0,V) = -\frac{\hbar\alpha(0)\omega_s}{16\pi z_0^3}\int_0^\infty \left[\frac{(1+\eta)\theta(1-\beta x)}{(1+\eta)^2 - \beta^2 x^2} + \frac{\theta(\beta x - 1)}{1-(\eta+\beta x)^2}\right][K_0(x) + K_2(x)]x^2\, dx \qquad (12)$$

where $\eta = \omega_s/\omega_0$, $\beta = V/2\omega_0 z_0$, $K_{0,2}(x)$ are the McDonald's functions and $\theta(x)$ is the unit step–function. Bearing in mind

$$\int_0^\infty (K_0(x)+K_2(x))x^2\, dx = 2\pi, \quad \int_0^\infty (K_0(x)+K_2(x))x^4\, dx = 12\pi,$$

one exactly retrieves (11) from (12) in the case $\beta \ll 1$. Integral (12) rapidly converges at any values of $\beta, \eta$.

In order to calculate $\Delta U(z_0,V)$, we employ for the imaginary part of the atomic polarizability the form

$$\alpha''(\omega) = \frac{\pi\alpha(0)\omega_0}{2}[\delta(\omega-\omega_0) - \delta(\omega+\omega_0)] \qquad (13)$$

Using eq.(13) and $\Delta'(\omega)$ corresponding to approximation (6), yields

$$\Delta U(z_0,V) = \frac{2\hbar}{\pi^2}\int_0^\infty dk_x \int_0^\infty dk_y\, k \exp(-2k z)\int_0^{k_x V} d\omega\, \Delta'(\omega)\alpha''(\omega - k_x V) = \\ = -\frac{\hbar\alpha(0)\omega_0\eta^2}{16\pi z^3}\int_{1/\beta}^\infty \frac{x^2(K_0(x)+K_2(x))}{\eta^2 - (1-\beta x)^2}\, dx \qquad (14)$$

It is obvious, that at $\beta \ll 1$ integral (14) quickly goes to zero and, therefore, the resulting van der Waals energy is given by eq.(11). A discontinuity point in the denominator of the integrand function (14) does not affect convergence of the integral. It completely vanishes if use is made of Drude –like form of $\varepsilon(\omega)$ with finite damping.

Also, it is of interest to get expressions for $U^{(0)}(z_0,V)$ and $\Delta U(z_0,V)$ in the case of a dielectric half-space, choosing $\varepsilon(\omega)$ to be

$$\varepsilon(\omega) = \varepsilon_\infty + \frac{(\varepsilon_0 - \varepsilon_\infty)\omega_T^2}{\omega_T^2 - \omega^2} \tag{15}$$

where $\omega_T$ is the transverse optical phonon frequency, $\varepsilon_0$ and $\varepsilon_\infty$ are the static ($\omega = 0$) and optical ($\omega \to \infty$) values of the dielectric permittivity. Substitution into (5) via (7),(13), (15) yields

$$U^{(0)}(z_0,V) = -\frac{\hbar \alpha(0)\omega_0}{16\pi z_0^3} \int_0^\infty [K_0(x) + K_2(x)]x^2 \cdot$$
$$\cdot \left\{ a_1\theta(1-\beta x) + \frac{d_1\eta(1+\eta)\theta(1-\beta x)}{[(1+\eta)^2 - \beta^2 x^2]} + \frac{d_1\eta\theta(\beta x - 1)}{[1-(\eta+\beta x)^2]} \right\} dx \tag{16}$$

$$\Delta U(z_0,V) = -\frac{\hbar \alpha(0)\omega_0}{16\pi z_0^3} \int_{1/\beta}^\infty [K_0(x) + K_2(x)]x^2 \cdot \left[ a_1 + \frac{d_1\eta^2}{\eta^2 - (1-\beta x)^2} \right] dx \tag{17}$$

$$a_1 = \frac{\varepsilon_\infty - 1}{\varepsilon_\infty + 1}, \quad b_1 = \frac{\varepsilon_0 - 1}{\varepsilon_\infty - 1}, \quad c_1 = \frac{\varepsilon_0 + 1}{\varepsilon_\infty + 1}, \quad d_1 = a_1 \frac{b_1 - c_1}{c_1} = \frac{2(\varepsilon_0 - \varepsilon_\infty)}{(\varepsilon_0 + 1)(\varepsilon_\infty + 1)} \tag{18}$$

## 4. Numerical examples

Now it is of great interest to illustrate the obtained eqs.(12), (14), (16) and (17) by making some numerical calculations. To do this, we have chosen parameters of metastable $He^*$ and ground state $He$ atoms, and of gold and silicon dioxide material parameters for the surface. The needed numerical values are listed in Table 1. Parameters of atoms were taken from [6] and of surfaces – from [7]. We have chosen $He^*$ and $He$ atoms in order to compare the cases $\eta > 1$ and $\eta < 1$ (see.eqs.(12), (14)) for walls made of metal.

Figs. 2(a,b) and 3(a,b) show relative values of the van der Waals energy with respect to its static values in dependence of parameter $\beta = V/2\omega_0 z_0$ for $He^*, He$ atoms interacting with gold (Fig.2) and $SiO_2$ (Fig.3) surfaces. The static van der Waals energy is given by eq.(11) at $V = 0$. Figs. 4,5(a,b) display the corresponding functions $U(z_0, V)$ given in absolute units vs. distance $z_0$ and velocity $V$. As can be seen from these figures, the dynamical van der Waals energy of a moving atom and a metal (dielectric) surface is essentially different as compared to the static case: starting from $\beta = 0$, the relative interaction energy (see solid lines on Figs.2,3) increases by 10 -20 %, and then goes to zero more abruptly with increasing parameter $\beta$. The positions of maxima depend on type of the atom and surface. In the case $He^* - Au$ (Fig.2(a)) one can see that at $\beta > 5$ the fraction $U^{(0)}(z_0, V)/U(z_0, 0)$ becomes even negative, corresponding to slight repulsion. However, as the contribution $\Delta U(z_0, V)$ proves to be much larger (it is shown by dotted line), the total interaction potential turns out to be attractive. This case corresponds to $\eta > 1$ in eq.(14). An opposite situation at $\eta < 1$ corresponds to $He - Au$ interaction (Fig.2(b)). In this case, in general, a small repulsive effect (however, not seen on Fig.2(b)) can be realized for $\Delta U(z_0, V)$, whereas the function $U^{(0)}(z_0, V)$ proves to be negative everywhere, giving rise to attraction. A minor distinction when the surface is taken to be dielectric (Fig.3), is that the maximum of $U^{(0)}(z_0, V)/U(z_0, 0)$ is shifted to smaller $\beta$ values. Figs.4,5 demonstrate large decreasing (by modulus) of the van der Waals attraction at distances of $0.3 \div 1 nm$. With increasing the distance and (or) decreasing velocity this effect becomes smaller and the dynamical potential asymptotically goes to the static one (see also Figs.2,3 at $\beta \to 0$).

## 4. Conclusions

For the first time, using nonrelativistic and nonretarded approximation of fluctuation electromagnetic theory, we have represented the dynamical van der Waals energy of a moving atom and the surface in the form of two terms, of which the second one asymptotically goes to zero when $V/2\omega_0 z_0 \to 0$, whereas the former tends to the static van der Waals energy with small velocity corrections ($\sim V^2$, to lowest order), when the velocity is small enough. Of principal significance is the possibility of representation of this part of interaction energy in the form of imaginary part of the integral over imaginary frequencies at any nonrelativistic velocities.

In general case, $V/2\omega_0 z_0 \sim 1$, both contributions to the van der Waals energy prove to be significant and should be taken into account, while the corresponding dependence on velocity and distance becomes strongly nonlinear.

The numerical calculations show that in the case of metal ($Au$) and dielectric ($SiO_2$) surfaces, the dynamical van der Waals interaction energy proves to be significantly smaller (by modulus) at velocities of $10^6 \div 10^7\ m/s$ and the distance to surface of order $0.3 \div 1 nm$. With increasing the distance and (or) decreasing velocity, the dynamical van der Waals potential tends to its static values more or less quickly in dependence of the wall material or type of atoms. We believe that these features are of great interest for experimental investigations in atom –surface scattering.

## Acknowledgements

We wish to express out great thanks to Prof. Gabriel Barton for inspiring comments and encouragement during writing this paper.

**Table 1**

**Parameters of surfaces and atoms**

| Surface | $\omega_p, \omega_T$ (eV) | $\varepsilon_0$ | $\varepsilon_\infty$ | Atom | $\alpha(0), 10^{-30} m^3$ | $\omega_0, eV$ |
|---|---|---|---|---|---|---|
| $Au$ | 9 | | | $He^*$ | 46.8 | 1.18 |
| $SiO_2$ | 0.328 | 4.88 | 1.0 | $He$ | 0.205 | 24.6 |

FIGURE CAPTIONS

Fig.1 Coordinate system used and geometry of atom –surface interaction.

Fig.2(a,b) Fraction of dynamical van der Waals energy (eqs.(5),(12), (14)) to the static one (eq.(11) at $V = 0$) for $He^*$ (a) and $He$ (b) atoms interacting with the surface of $Au$ in dependence of parameter $\beta = V/2\omega_0 z_0$. Dashed line –eq.(12); dotted line –eq.(14); solid line – sum of (12), (14).

Fig.3(a,b) The same as on fig.2 in the case of $SiO_2$ surface.

Fig.4(a,b) Resulting dependence $U(z_0,V)$ vs. distance $z_0$ and velocity $V/V_B$ ($V_B = 2.2 \cdot 10^6 \, m/s$ is the Bohr velocity) for $He^*$ (a) and $He$ (b) atoms interacting with $Au$ surface. Solid line on Fig.4(a) corresponds to $V/V_B = 5$, dotted line –to $V/V_B = 2$; dashed line –to $V/V_B = 1$; and dashed –dotted line –to $V/V_B = 0.5$. The same lines on Fig.4(b) correspond to $V/V_B = 15, 10, 5, 1$, respectively.

Fig.5(a,b) The same as on fig.4 for $He^* - SiO_2$ (a) and $He - SiO_2$ (b). Solid, dotted, dashed and dashed –dotted lines on Fig.5(a) correspond to $V/V_B = 5, 2, 1, 0.5$ and $V/V_B = 15, 10, 5, 1$ on Fig.5(b).

FIGURE 1

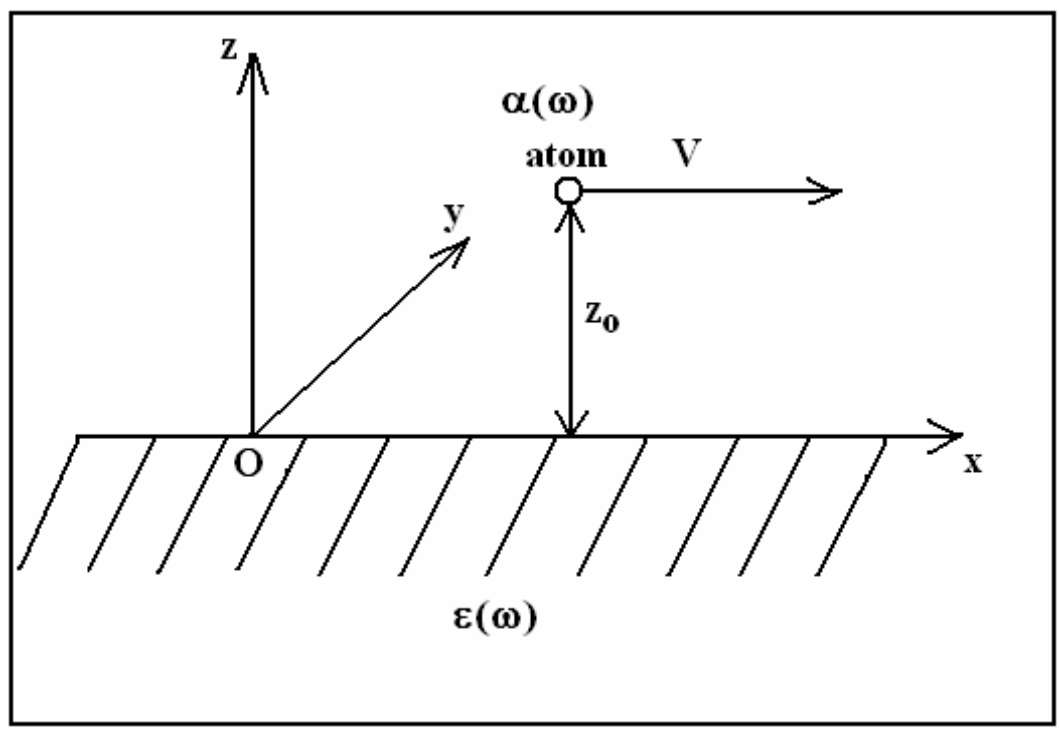

FIGURE 2

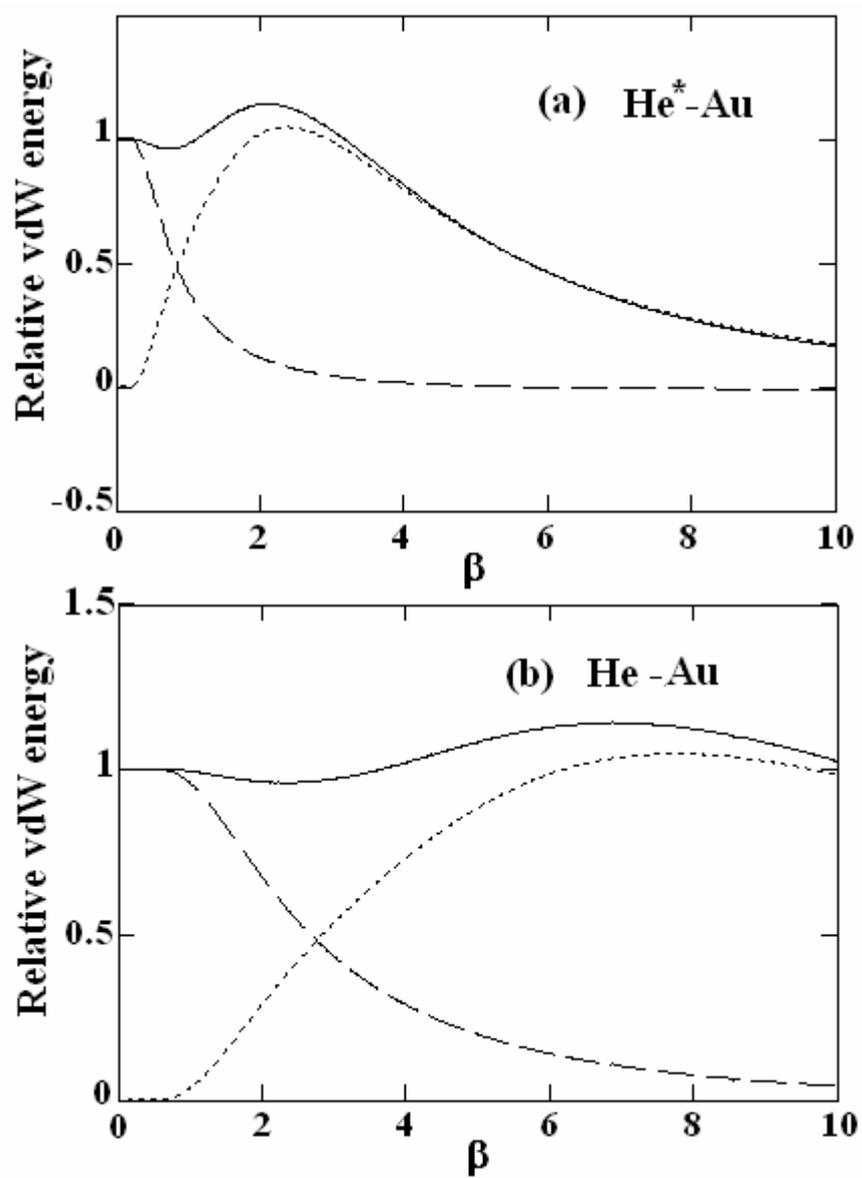

FIGURE 3

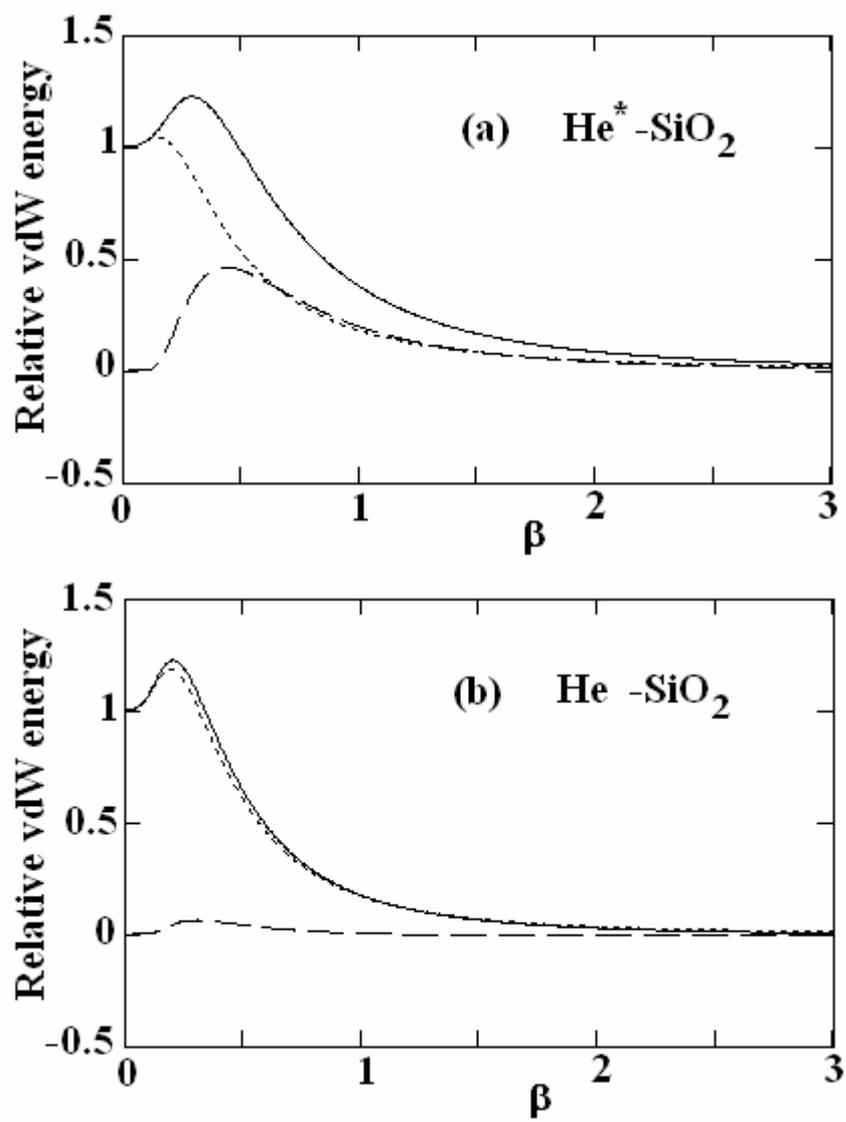

FIGURE 4

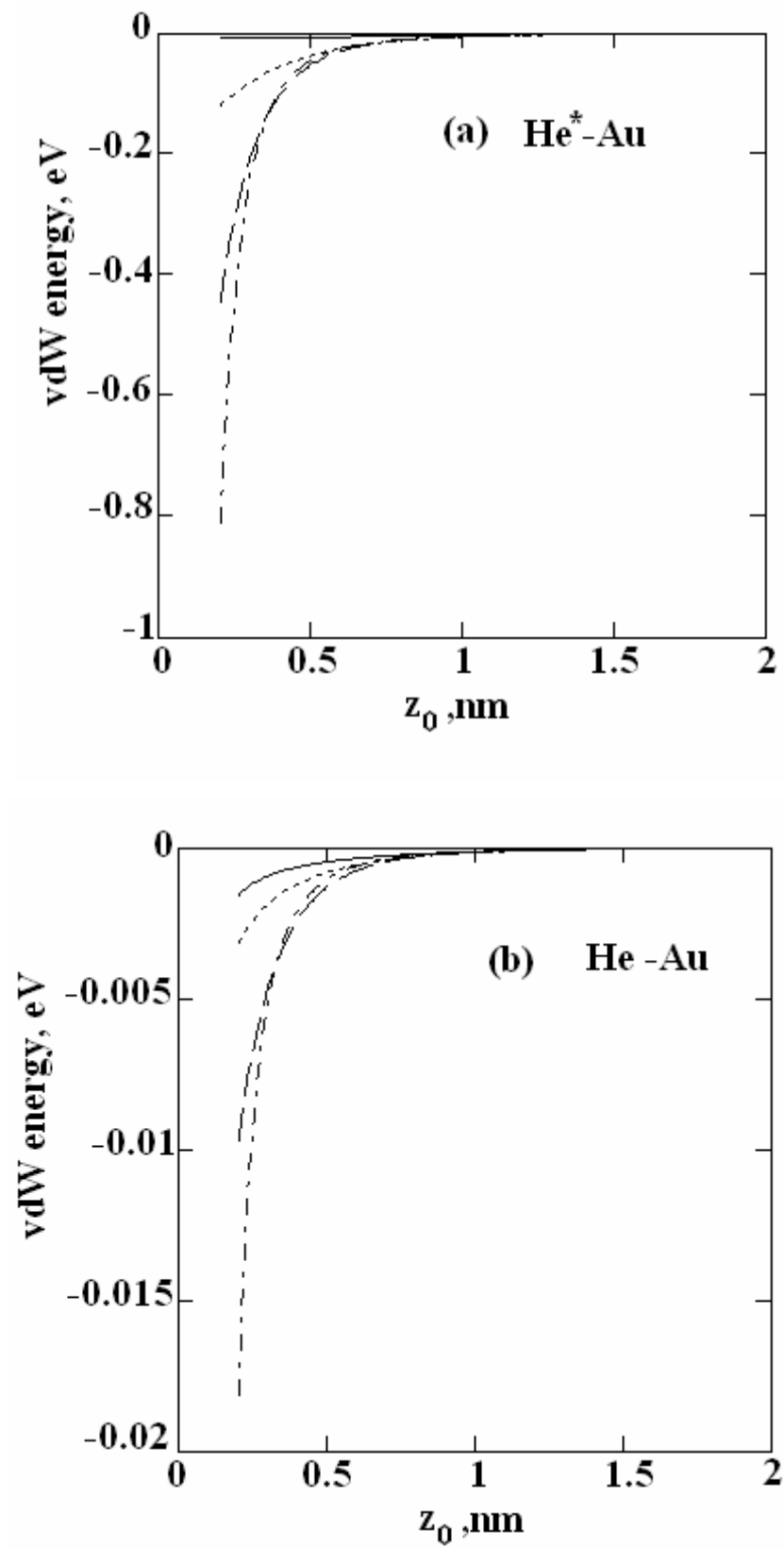

FIGURE 5

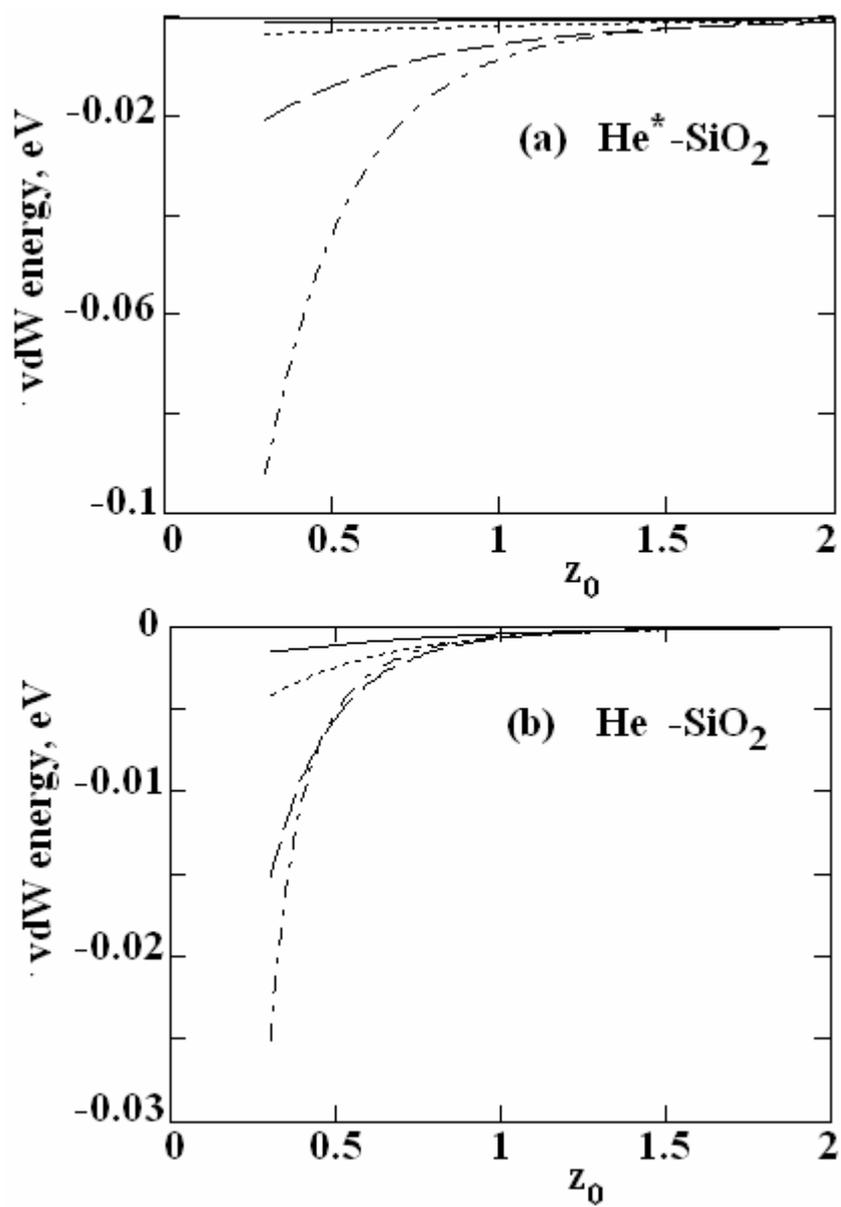